# Generative AI-Enabled Adaptive Learning Platform: How I Can Help You Pass Your Driving Test?

Riya Gill[1], Ievgeniia Kuzminykh[1*], Maher Salem[1], Bogdan Ghita[2]

[1]Department of Informatics, King's College London, Strand, London, WC2R 2LS, United Kingdom.
[2]School of Engineering, Computing, and Mathematics, University of Plymouth, Drake Circus, Plymouth, PL4 8AA, United Kingdom.
*Corresponding author(s). E-mail(s): ievgeniia.kuzminykh@kcl.ac.uk

**Structured abstract**
**Purpose**
This study aims to develop an adaptive learning platform that leverages generative AI to automate assessment creation and feedback delivery. The platform provides self-correcting tests and personalised feedback that adapts to each learner's progress and history, ensuring a tailored learning experience.

**Design/methodology/approach**
The study involves the development and evaluation of a web-based application for revision for the UK Driving Theory Test. The platform generates dynamic, non-repetitive question sets and offers adaptive feedback based on user performance over time. The effectiveness of AI-generated assessments and feedback is evaluated through expert review and model analysis.

**Findings**
The results show the successful generation of relevant and accurate questions, alongside positive and helpful feedback. The personalised test generation closely aligns with expert-created assessments, demonstrating the reliability of the system. These findings suggest that generative AI can enhance learning outcomes by adapting to individual student needs and offering tailored support.

**Originality**
This research introduces an AI-powered assessment and feedback system that goes beyond traditional solutions by incorporating automation and adaptive learning. The non-memoryless feedback mechanism ensures that student history and performance inform future assessments, making the learning process more effective and individualised. This contrasts with conventional systems that provide static, one-time feedback without considering past progress.

## 1. Introduction

The integration of artificial intelligence (AI) into education has transformed traditional learning paradigms, enabling more personalised, accessible, and effective educational experiences. In recent years, the emergence of generative AI (GenAI) models such as ChatGPT, Claude, and Gemini has further expanded possibilities for adaptive learning. However, despite their potential, these tools remain underutilised in structured educational settings, with research predominantly focused on ethical considerations, policy frameworks, and AI literacy rather than holistic, automated learning solutions (Pesovski *et al.*, 2024).

Traditional education often adopts a structured, teacher-centred approach, relying on standardised materials that prioritise memorisation over deep comprehension (Gowda and Suma, 2017). While sufficient for conventional testing, this model fails to accommodate individual learning styles, leaving struggling students disengaged and demotivated (FairTest, 2012). In response to these gaps, students have begun to use new mainstream GenAI tools such as ChatGPT and Grammarly (Nagelhout, 2024). Although popular and widely accessible, this use is unregulated on the students' side, as well as potentially ineffective, as AI-generated content can contain misinformation or bias (Monteith *et al.*, 2024) and can unintentionally hinder the learning and development of the student. Educators need to stay ahead of this technology.

There are several custom solutions (usually in the form of GitHub projects) that connect learning management systems like Moodle to ChatGPT, for example, the GitHub Moodle plugin (Yoder, n.d.) that automatically generates questions from the text. In addition, research papers evidence that there are custom solutions in different education institutions that develop tools that use generative AI for grading the assessments and providing feedback (Kuzminykh *et al.*, 2024; Blommerde *et al.*, 2024). However, they are limited to institutional usage and can't go beyond due to ethical, GPDR issues and data protection. Moreover, none of them has an adaptability component when every next generated set of questions or assessment considers the previous performance of the student or their answer score.

This study aims to develop a personalised adaptive learning platform which uses generative AI to create a unique learning experience based on each learner's needs, preferences, goals, and progress. Generative AI will be used to develop customised learning materials such as questions and tests, and additionally, individualised feedback will be provided to allow for a deeper understanding of the material. Individualised feedback will be provided in the form of graphs tracking the learner's progress, highlighting weaker areas. To facilitate the customised content, the application will continually track the learner's progress and adapt, based on the learner's development over time. allowing users to set goals and focus on weak areas. The learner should still be able to control their own learning through the ability to set their own goals or choose the specific topic they wish to learn. This platform will be evaluated to help assess the benefits of GenAI for education.

This study focuses exclusively on text-based UK Driving Theory Test preparation for learners aged 17+, excluding multimedia content and practical assessments. The platform's effectiveness will be measured by its ability to deliver personalised learning experiences through adaptive question generation and performance tracking. As theory test preparation is typically self-taught, the system operates without educator interfaces, functioning as an autonomous learning tool.

The rest of the paper is structured as follows: Section 2 reviews the current landscape of GenAI, explores the integration of GenAI with learning, and evaluates both the benefits and challenges it presents to learners. In Section 3, the method and tools behind the adaptive platform are explained. The implementation and results are presented in Section 4, followed by the discussion of the results in Section 5. Finally, we conclude the work in Section 6.

## 2. Related Studies

### 2.1. Ethical and Pedagogical Concerns

As related studies show, generative AI can enhance teaching and learning practices through automatic content creation, learning design and immediate feedback. Gianakos *et al.* (2024) discuss these benefits while also acknowledging the challenges "hastily adopting GenAI tools" can have on the educational system. The article argues that while tools like this are surrounded with questions on ethics and environmental concerns, it is vital we do not abandon them and instead identify further avenues for research and provide the necessary framework of legislation, policy and support to facilitate and utilise these tools in ways that can be revolutionary.

The hesitation surrounding GenAI adoption is not without merit. Experts emphasise the need for ethical deployment to mitigate risks such as reduced critical thinking skills. Alasadi and Baiz (2023), and Qadir (2023) observed that overreliance on GenAI tools can hinder the ability of the students to analyse and synthesise information independently. Qadir (2023) notably states, "It is more important to ask good questions than to learn how to answer them", highlighting the risk of passive learning. However, Quadir's work is primarily theoretical, lacking empirical case studies to substantiate and validate these claims.

Legal and ethical issues further complicate GenAI's use. Large language models (LLMs) such as ChatGPT scrape unverified data from the internet, often reproducing biased or plagiarised content (Monteith *et al.*, 2024; Haiqiong and Hoiio, 2024). Monteith *et al.* (2024) stress that, without proper citation mechanisms, GenAI tools risk normalizing intellectual property violations. In the study by Haiqiong and Hoiio (2024) on ChatGPT plagiarism, the research methodology used is limited to analysing a number of Chinese research papers, with the focus being on only rewriting sources and does not mention other forms of academic work. Potentially studying other forms of use of AI may provide more specific examples of how to educate students and provide a more comprehensive understanding of AI's influence in education.

Many educators have discussed the implications of GenAI tools and provided their insight into the specific pitfalls GenAI has within the education field. Ifenthaler *et al.* (2024), as part of their academic staff survey, found that privacy, data security, and algorithmic transparency were top concerns. Participants feared that unchecked GenAI autonomy might lead to unreliable or unpredictable outcomes. To address this, Ifenthaler *et al.* advocate in the same study for enhanced AI literacy among educators and robust regulatory policies. While their investigation does provide valuable insight into AI trends, it has a limited sample size with potential biases in expert opinion. Future research should look to incorporate more diverse participants to provide more robust findings.

### 2.2. Adaptive Learning Platforms

Adaptive learning platforms (ALP) specifically are potentially revolutionary within the education technology sector. These systems tailor content to individual learners needs using data-driven algorithms and offering real-time feedback and resource efficiency [14]. Further, with the current capabilities of GenAI content creation in various mediums such as video, images and text, AI has the potential to aid learners who require varied learning approaches and diverse learning needs. In addition, such a platform can dynamically adjust lesson difficulty or reuse materials, reducing teachers' administrative burdens.

However, Esan (2024) identifies significant drawbacks, arguing that the lack of accountability and transparency can have a more substantial impact, as users are inherently trusting the system to teach them. Learners go into the environment with an open learning mindset, making them potentially less critical of information being presented to them. These systems can then influence the learner. ALPs also may strip the learners ability for self-regulation as they provide users with information about how they are performing, rather than the learner self-regulating their performance. This may result in users becoming unable to develop critical self-understanding skills and could consequently develop a dependence on these AI systems. Esan, while providing initiative ideas, may benefit from real-world examples to provide more credibility to this claim.

*2.3. Student Focus*

Students are among GenAI's most active adopters, using tools like ChatGPT for research, writing assistance, and exam preparation. Surveys indicate that banning GenAI is impractical, as 60–70% of university students already use it for academic tasks (Johnston *et al.,* 2024). Chan and Hu (2023) focus on the current learner's perspective of GenAI. They highlight its utility for non-native English speakers, who leverage text-to-text generators to improve writing fluency. GenAI also aids research by summarising complex materials and identifying key patterns.

Chan and Hu (2023) also note that using tools such as ChatGPT to mark essays with standardised scoring as a consistent marking criterion. This helps normalising the evaluation system and removes human bias. It also provides the ability to formatively guide the students through the process without taking over the learning. At the other end, the students' perceptions and efficient use of tools like GenAI stem from their initial attitudes towards learning and their environments. Students with positive associations with learning and those who have confidence in their abilities would conduct more thorough and in-depth research and develop a deeper understanding of the material being taught, using ChatGPT as an additional tool rather than as an alternative for their learning effort. The reverse of this is also true, whereby students who do not have confidence in their abilities or have a negative perception towards learning conduct surface-level research and passively consume information. Chan and Hu argue that these would affect the integration of AI into learning as there is variable willingness to use these tools.

There is currently limited research into the use of GenAI in learning, but many studies do find that AI is useful due to its interactivity, responsiveness, and support, and can bring benefits such as enhanced learning experiences, increased motivation, and positive career impacts. Bisdas et. al. explored the frequency of AI use and student perceptions and argue that there is a positive association between learner AI use and confidence gained (Bisdas *et al.,* 2021). Other studies, such as Durak (2023), suggest there is no meaningful connection between the two.

*2.4 Existing AI-powered Learning Platforms*

Pesovski *et al.* (2024) explore the integration of GenAI and personalised learning using a tool in the context of a software engineering college, offering personalised quizzes and multi-format content (e.g., videos, text). The approach was evaluated through a study using 20 students to assess the impact of the tools used. Findings indicate the students enjoyed the variety of learning formats available and primarily used the traditional materials. Students also reported that the quizzes were particularly valuable as they aided in deeper understanding and retention of learning materials. The study also reported increased study time, particularly with students who had initially struggled with the topics. Notably, the tool had no negative effects, suggesting well-designed GenAI can complement traditional teaching. The authors therefore concluded that GenAI has a positive impact in education by enabling personalised learning experiences to enhance student engagement and learning. Their results showed no negative implications for students using GenAI technologies and only reported a positive effect for the students who were initially underperforming. In relation to these points, the analysis demonstrated that the students using the technology were those who were underperforming, as they are typically not catered for with the current 'one-size fits all' form of education. The ability to have different methods of content delivered to these students allows them to have potential content which fits their needs, or content which is more thorough and explained. They conclude that the implementation of GenAI in teaching and learning should enable these personalised benefits. The paper ultimately critically analyses the full process of utilising AI-powered content generation in an educational setting, from the perspective of its impacts.

A large-scale example of an AI-assisting education system is China's Squirrel AI, an adaptive tutoring platform used by millions (Hao, 2019). It begins with evaluation tests to create customised curricula, breaking subjects into granular "knowledge points" for targeted instruction. These refined points allow the system to target specific aspects of the learner knowledge gaps and to address them, Hao (2019) does mention that educational experts suggest that Squirrel AI does not fully align with the skills needed for the modern era, highlighting the difference between "adaptive" and "personalised" learning. In this context, "adaptive learning" seeks to solely identify what the student knows and does not know, while personalised learning aims to understand what the student wants to learn, taking their needs and interests into account in order to "orchestrate the motivation and time for each student so they are able to make progress". The former lacks creativity, collaboration, or the ability to learn autonomously, all vital skills for a successful learning journey.

*2.5 AI in Non-academic Context*

The research reviewed so far focuses primarily on the academic learning process; this is justifiable as primary, secondary, and higher formal education are the typical examples for learning environments, hence the direct application areas for any AI-based solutions. However, when investigating learning journeys, one should also consider non-standard areas, such as the theory driving test. While the ecosystem of the theory test for the driving license may differ in terms of body of knowledge or depth, it does raise interesting challenges due to the target student audience, implications for knowledge, and approaches to study. To begin with, unlike the focused learning paths mentioned, the student audience is more diverse in terms of prior knowledge, demographics, expertise, and highest level of study attained; this requires a more inclusive approach when developing the questions and associated impact, to consider the diversity of the audience. The approach to study may also vary within the student population, from segments that are familiar with and still going through a form of education to others that may not have had any education or limited exposure to learning in recent years. Finally, while the purpose of the learning is indeed to pass the theory exam, the knowledge acquired is likely to be required in critical decision-making situations, with no access to sources of information, where timing is vital.

In this context, the UK Driver and Vehicle Standards Agency (DVSA) acknowledged in 2025 that it has used AI for "creating theory test questions prior to them being tested in a controlled test environment" (UK Parliament, 2025).

The driving industry, together with the learning community, is also aiming to leverage AI to ensure optimal driver responses and reactions; a typical example is the work by Murtaza et al. (2023) where the authors explore the integration of ADAS (Advanced Driver Assistance Systems) and Autonomous Vehicles (AV) with an AI assistant to deliver a more coordinated approach to driving. Their work stems from the wider domain of assistance, such as Yang *et al.* (2021) that considers the environment of pilots and flight simulators, and paves the way towards. AI monitoring of driver behaviour can also be used as a medical diagnosis tool, as successfully demonstrated by Al-Hindawi *et al.* (2025).

While AI is being employed in various areas relating to driving, the actual theory test remains outside the research area; currently there are commercial companies advertising AI-based learning, with claims such as Driving Theory 4 All (2024) with "AI-powered theory tests", but with no clear description of the AI assisting element.

To conclude, the learner and driving area does have a number of avenues that do benefit from the embedding of AI, but so far, no study has considered a scientific approach or review within this context.

*2.6 Research Gap*

GenAI offers transformative potential for education through personalised learning and rapid feedback. It also presents challenges that include ethical concerns and an over-reliance on AI use that may lead to a lack of critical thinking and self-regulation skills, as highlighted by many of the reviewed research papers. To mitigate these challenges, the aim of this study is to propose a platform that will focus on knowledge examination using multiple-choice questions based on specific content. This approach ensures that the role of AI is primarily to provide feedback and assess knowledge rather than assist in answering the questions directly. Given the challenges raised by prior studies, this approach will allow learners to engage more deeply with the material and develop critical thinking skills rather than become information consumers. Adaptive learning platforms can encourage diverse learner progression and needs, potentially helping to encourage underachieving learners. There is a noticeable lack of papers mentioning other forms of learning that are not school-based education (such as training courses, vocational learning, or adult education). Further research is needed into these subjects across all areas of education to maximise the benefits that can be achieved and ensure all downsides are accounted for. Our study seeks to help address this issue by focusing on self-administered testing for learners aged 17+ and above, such as those preparing for driving theory tests or other adult-focused assessments.

The related papers are categorised by content type, mentioned models, student level, feedback provided by AI, learning context, the authors of the paper, and whether progress tracking and presented in Table I.

Table I. Overview of related studies

| Study | Type | Model | Level | Feedback | Context | Progress Tracking |
|---|---|---|---|---|---|---|
| (Adams *et al.*, 2006) | Application and comparison | E-learning applications | University | Yes | Education | Yes |
| FairTest, 2012) | Ethics and research implementation | N/A | Primary-secondary education | N/A | Educational | No |
| Nagelhout, 2024 | Research exploration | General GenAI models | Teenagers – young adults | No | Education, general use of GenAI | No |
| Monteith *et al.*, 2024 | Ethics | ChatGPT, GPT-3, GPT-4 | General Public, Medical/Healthcare Professionals, All level of students | N/A | Medical Information | N/A |
| Giannakos *et al.*, 2024 | Research, Ethics, Feedback | ChatGPT, LLMs, OpenAI Codex | University-level | Yes | General education, programming, law, medicine and auditing | Yes |

| Study | Type | Model | Level | Feedback | Context | Progress Tracking |
|---|---|---|---|---|---|---|
| Skelton, 2022 | Ethics and policy discussion | N/A | Academic, Industry | No | N/A | N/A |
| Alasadi and Baiz, 2023 | Practical applications of AI | GPT-4 (ChatGPT) | University-level | Yes | N/A | N/A |
| Qadir, 2023 | Ethics | ChatGPT | Traditional education | Yes | General education | No |
| Haiqiong and Hoiio, 2024 | Ethics | ChatGPT | University | N/A | Academic | N/A |
| Ifenthaler *et al.*, 2024 | Policy and research trends | N/A | General | N/A | Writing | N/A |
| Esan, 2024 | Research implementation | N/A | General | Yes | General education | Yes |
| Johnston *et al.*, 2024 | Research into students | ChatGPT, Grammarly | University | No | Academic writing | N/A |
| Chan and Hu, 2023 | Research into students | ChatGPT and other AI tools | University | N/A | Academic support | No |
| Bisdas *et al.*, 2021 | Students' perspective | N/A | University | N/A | Medicine and dentistry education | N/A |
| Durak, 2023 | Research implementation | General chatbot | University | N/A | Visual design, Software Engineering, General education | N/A |
| Pesovski *et al.* 2024 | Research implementation | OpenAI | University | Yes | N/A | Yes |
| Hao, 2019 | Research implementation | Squired AI's Intelligent Adaptive Learning System | Primary-Secondary education | Yes | N/A | Yes |

## 3. Methodology

To achieve the aim of the study, the AI-driven learning driving theory platform was implemented in a web application that provides a learner with questions contained within quizzes, accompanied by statistics on their performance. The quizzes contain GenAI questions, generated to suit the specific learning needs of the user. The platform, shown in Fig.1, uses a standard architecture, consisting of a front-end web application, a database, and GenAI model responsible for dynamically generating the personalised learning content. This block uses prompt engineering to create customised questions, explanations, and feedback based on the learner progress. It ensures that content is adaptive and aligned with the user needs while maintaining accuracy and coherence.

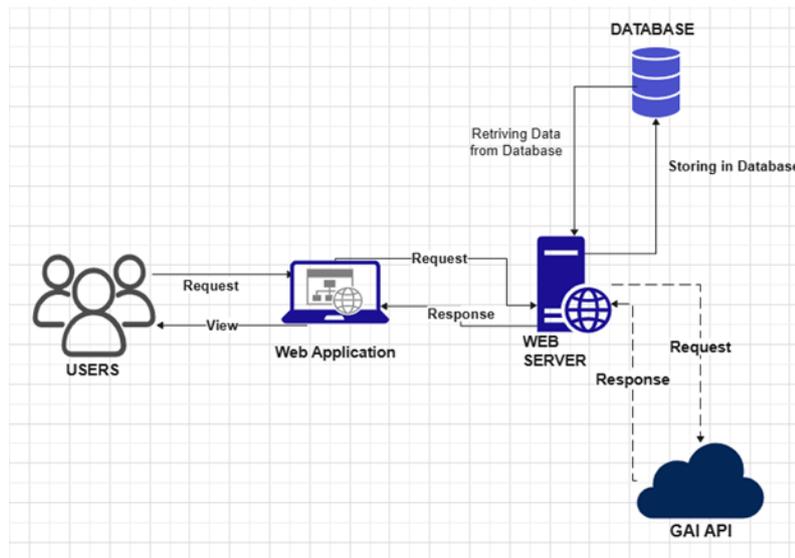

Fig.1. The architecture of the adaptive learning platform displaying communication between components of the web application

*3.1 LLM Model Selection*

The web application uses GenAI via the Google Gemini Flash model (Google, 2024) accessed through the Google Cloud Console. Gemini Flash was chosen for its integration of multiple machine learning models, ease of use, flexibility, and "impressive speed" (Quintero et al., 2024). Two interaction methods were considered: prompt engineering and fine-tuning. Although fine-tuning with UK Driving Theory Test questions was initially planned, the limited availability of open-source data and the impracticality of collecting sufficient material led to the adoption of prompt engineering. This method enables diverse AI-generated content with simpler implementation (Amazon Web Services, 2025).

To further enhance the model's performance, a Retrieval-Augmented Generation (RAG) system retrieves user-specific data from the database, reducing hallucinations and improving contextual relevance. While embedding-based retrieval could offer even greater efficiency by structuring and storing knowledge, its computational demands make it unsuitable for the current lightweight web application but worth considering for future development.

*3.2 Evaluation and Performance Metrics*

The application has three main requirements in relation to GenAI: dynamic question generation, personalised feedback, and topic distribution.

To assess the accuracy of the GenAI model, both model-on-model evaluation and expert evaluation were applied to feedback and test questions, which are content generated by generative AI and are based on (Das *et al., 2023*). The model-on-model evaluation involved assessing the generated questions using an additional model, the more advanced Gemini-2.0-Pro-Exp-02-05 model. The expert evaluation used a focus group of eight people who passed their UK Driving Theory Test within the past three months or are in the process of preparing for the test. The participants were undergraduate university students who were required to be users of existing mobile applications for UK Driving Theory Test preparation, so they can provide supportive insights into the validity and relevance of the generated material.

For generated questions, a set of five tests was conducted, each including ten topic-based tests. All generated questions were assessed based on the criteria presented in Table II.

Table II. Criteria for question evaluation

| Percentage | Criteria | Question asked | Explanation |
|---|---|---|---|
| 40% | Answer Accuracy | Are the provided answers correct? Are they valid compared to the other answer choices? | Check no hallucinations are occurring and information is factually correct. |
| 40% | Question Relevancy | Do the generated questions align with the expected topic? | Ensures questions are relevant and are not misleading, so users are tested on the correct material. |
| 20% | Diversity Value | How similar are the questions? | Assesses whether the generated questions go beyond traditional questions and promote deeper understanding. |

Both generated feedback for specific questions and overall test feedback were evaluated using the rubrics presented in Table III. Similar methods of evaluation were seen in the work of Steiss *et al.* (2024) and Kuzminykh *et al.* (2024), where they compared GPT-generated feedback for free-text questions against teacher-given feedback.

Table III. Criteria for feedback evaluation

| Percentage | Criteria | Question asked | Explanation | Notes |
|---|---|---|---|---|
| 40% | Accuracy | Is the feedback content accurate and true to the information it is based on? | Verifies feedback is factually correct, so as not to mislead the user. | Metric for both overall and per question feedback |
| 40% | Personalisation and guidance | Is the feedback helpful, clear explain the concept or error correction and does it provide the user with information on how to proceed? | Assesses whether the feedback provides actionable insights which guide the user on how to improve. | Metric for both overall and per question feedback |
| 20% | Relevancy | Does the feedback make sense in relation to the information it is referring to? | This ensures that feedback relates to the question, avoiding vague or off-topic responses that could confuse the user. | Only for feedback to question |
| 20% | Positivity | Is the test feedback encouraging and uplifting? | Evaluates whether the feedback is encouraging and supportive, helping the user maintain motivation and engagement. | Only for overall feedback |

As part of the evaluation, side-by-side mean error comparison was used to determine the average error between expected and actual values and to benchmark the personalisation of the mock test according to the user's needs.

### 4. Results and Evaluation

*4.1 Model Performance for Question Generation*

The questions were evaluated on three performance metrics from Table II: relevancy, accuracy, and diversity. Relevancy and accuracy were based on the categories **strong yes**, **weak yes**, **weak no**, **strong no**, whereas diversity is ranked on a 1 to 5 scale where 1 is 'very similar' and 5 is 'completely different'.

For the first two criteria, we used 100 questions generated by our application in topic-based tests. Then these questions were passed to both the model (Gemini-2.0-Pro-Exp-02-05) and the experts.

For example, say the application generates the question "What is the national speed limit for a UK motorway?" with the correct answer "70mph" and other options including "60mph", "80mph", and "50mph". The prompt for the Gemini model used for relevancy evaluation will look as presented in Fig. 2.

```
The question generated is: '{question}'. With correct answer:
'{correct answer}' and additional option: '{additional answers}'. The
topic expected is '{topic}'. Does the generated question align with the
expected topic? Please answer with only the categories strong yes, weak
yes, weak no, strong no.
```

Fig. 2. Prompt used for question relevance assessment.

The diversity score was evaluated with real UK Driving Theory Test questions that have already been categorised and labelled using two categories ('Rules of the road' and 'Safety and your vehicle') with 50 model-generated questions. Questions diversity ratings were given by both a model and an expert. Fig. 3 shows the Gemini model prompt. Diversity refers to how original the question is; this could be differences in wording, scenarios or new questions which have not been asked. For benchmarking, ten authentic questions from each relevant topic were sourced from the "Theory 4 in 1" mobile application (Focus Multimedia Ltd, 2024). Each model-generated question was then compared against this set of reference questions.

```
Here is a generated question for the UK driving theory test:
{Generated Question}

Below are five related questions from the mobile app 'Theory 4 in 1':
{Reference Question 1}
{Reference Question 2}
{Reference Question 3}
{Reference Question 4}
{Reference Question 5}
On a scale of 1 to 5 (where 1 = very similar and 5 = completely different),
rate how similar the generated question is to each reference question.
```

Fig. 3. Prompt used for question diversity assessment.

A summary of the results is shown in Table IV.

Table IV. Performance evaluation for question generation

| Ranking | Accuracy | | Relevancy | | Ranking | Diversity | |
| --- | --- | --- | --- | --- | --- | --- | --- |
| | model | expert | model | expert | | model | expert |
| strong yes | 98 | 94 | 70 | 58 | 1 | 7 | 15 |
| weak yes | 2 | 6 | 27 | 33 | 2 | 35 | 47 |
| weak no | 0 | 0 | 3 | 9 | 3 | 30 | 21 |
| strong no | 0 | 0 | 0 | 0 | 4 | 12 | 9 |
| | | | | | 5 | 16 | 8 |

We performed statistical analysis to evaluate the agreement between model and expert assessments. The Pearson correlation coefficient for accuracy and relevancy showed a value of **0.96** ($p < 0.174$) and 0.99 ($p < 0.096$), respectively; the Chi-Square test showed $\chi^2 = 9.35$ and p ≈ 0.053. Both tests indicate a strong correlation between the model and the expert in terms of the relative ranking of accuracy and relevancy, showing they generally agree on which questions more or less reflect the content requested; *p-value* is not below the typical

0.05 threshold for significance due to the very small sample size. The Chi-Square test shows a marginal difference in score distribution, indicating the model may slightly under- or overestimate diversity compared to the expert.

In terms of diversity, Cohen's Kappa coefficient was used to measure inter-rater (out of two raters) reliability while accounting for agreement by chance. The value of κ = 0.10 showed slight agreement due to mismatched counts. The Pearson coefficient for diversity is r = 0.12 (p-value = 0.230), which indicates very weak alignment between the model and the expert, and the model systematically overestimates diversity compared to the expert. The expert rates more questions as low-diversity (Rank 1–2), while the model favours higher ranks.

The model and expert average for each metric, as well as the overall average performance, are shown in Table V.

Table V. Averaged performance of the question generation

|  | Accuracy (40%) | Relevancy (40%) | Diversity (20%) | Diversity (1-5) | Overall |
|---|---|---|---|---|---|
| Model | 0.993 | 0.88525 | 0.5125 | 2.95 | 0.8538 |
| Expert | 0.979 | 0.82375 | 0.63 | 2.48 | 0.8471 |

The results show that the model slightly outperforms the expert by 0.67 percentage points on a weighted average basis. The model's average diversity score (2.95) is higher than the expert's (2.48), indicating the model perceives questions as more diverse. The expert's lower score suggests they find questions less original overall, and they are stricter seeing more questions as similar (Ranks 1–2).

**Discussion:**

The evaluations demonstrate strong alignment between model-generated and expert in the assessing the question relevancy, though slight discrepancies emerge. The expert evaluation shows a more critical distribution toward 'weak yes' responses, suggesting this could be due to the expert's inherent subjective bias, domain expertise, or more rigorous evaluation criteria (Abeysinghe and Circi, 2024). The model's evaluation seems more lenient, as shown in the smaller number of allocations to 'weak no' and 'strong no'. This may imply that while the model is effective in generating relevant questions, there could be room for improvement in enhancing the slight nuances of relevancy that currently only the expert might identify.

Accuracy assessments consistently reveal correct answers, with "weak yes" ratings primarily reflecting ambiguous cases where multiple interpretations are possible; for instance, another answer could be seen as correct. This indicates that while the model performs well overall, there may be edge cases where questions can be interpreted in various ways, leading to discrepancies in the answer choices. Possible solutions to enhance question clarity include refining the training data or adjusting the prompt to provide clearer instructions for generating unambiguous questions and answers.

For diversity, both evaluation methods' average value lies between 2 and 3, suggesting questions are moderately diverse. This is understandable as LLMs work by using a large set of data to generate responses, so questions generated contain patterns from multiple sources. There are discrepancies at the extreme ends of the evaluations. The expert seems to view the diversity of the questions more critically, potentially identifying more subtle variations in content or structure. The expert's bias towards rating questions low suggests that they expect higher differentiation in terms of content, making them more critical of perceived repetition in the model's questions.

Overall, the generated questions performed well in terms of diversity, but there is still room for improvement in identifying areas where the questions could be adjusted to differentiate in a way the expert expects. This aligns with other results as LLMs tend to fall short in terms of originality (Zhao et al., 2024).

Questions performed well on the criteria discussed, and based on this it can be concluded that the current GenAI is able to produce high-quality questions for the application and can be a valuable tool for general question generation. To enhance its effectiveness, fine-tuning is recommended to ensure better alignment with experts' expectations. With this adjustment, the GenAI can be more effective and reliable in a real-world environment.

*4.2 Model Performance for Feedback*

Both question-specific and overall feedback to the user were evaluated based on the metrics from Table III. We expect that the question-specific feedback should be neutral and explain the correct answer. The overall feedback is based on the questions incorrectly answered by a user and should be personal and helpful.

Question-specific feedback was assessed on the following criteria: relevance to a question, whether the feedback is helpful and whether the feedback is clear and accurate. Overall test feedback was evaluated by the following criteria: positivity, helpfulness and accuracy of data. These were all rated by the categories 'strong yes', 'weak yes', 'weak no', 'strong no' for 50 pieces of feedback collected from the application. These were formatted into a prompt and passed to both the model and expert, with their response collected.

Complete results for the question-specific feedback and overall feedback are displayed in Table VI and Table VII, respectively.

Table VI. Performance evaluation for question-specific feedback generation

| Ranking | Accuracy | | Personalisation | | Relevancy | |
|---|---|---|---|---|---|---|
| | model | expert | model | expert | model | expert |
| strong yes | 48 | 48 | 48 | 49 | 48 | 47 |
| weak yes | 2 | 2 | 2 | 1 | 2 | 3 |
| weak no | 0 | 0 | 0 | 0 | 0 | 0 |
| strong no | 0 | 0 | 0 | 0 | 0 | 0 |

Table VII. Performance evaluation for overall feedback generation

| Ranking | Accuracy | | Personalisation | | Positivity | |
|---|---|---|---|---|---|---|
| | model | expert | model | expert | model | expert |
| strong yes | 48 | 45 | 47 | 46 | 44 | 44 |
| weak yes | 2 | 5 | 3 | 4 | 6 | 6 |
| weak no | 0 | 0 | 0 | 0 | 0 | 0 |
| strong no | 0 | 0 | 0 | 0 | 0 | 0 |

The averaged results presented in Table VIII demonstrate that 98-99% of feedback was regarded as positive, helpful, and accurate according to both evaluation methods. There is a negligible difference between model and expert evaluations, suggesting a shared understanding of the feedback's nature in both cases. This data indicates that question-based feedback is highly reliable and beneficial for educational purposes. Although there is potential for a slight increase in helpfulness, the responses are pertinent and accurate, implying that utilising this form of feedback alongside additional learning may be advantageous.

Table VIII. Averaged performance of the question generation

| | Question-specific feedback | | | |
|---|---|---|---|---|
| | Accuracy (40%) | Personalisation (40%) | Relevancy (20%) | Overall |
| Model | 0.986 | 0.986 | 0.99 | 0.9868 |
| Expert | 0.986 | 0.993 | 0.985 | 0.9886 |
| | Overall feedback | | | |
| | Accuracy (40%) | Personalisation (40%) | Positivity (20%) | Overall |
| Model | 0.986 | 0.979 | 0.97 | 0.98 |
| Expert | 0.965 | 0.972 | 0.97 | 0.9688 |

GenAI-generated feedback received overwhelmingly positive ratings in terms of the criteria stated. There were no 'weak no' or 'strong no' ratings indicating feedback is generally beneficial. The 'weak yes' on positivity was in relation to the test where a '0' score was achieved, it is understandable that in cases where all questions have been answered incorrectly, it is perceived as a response that is somewhat positive.

In general, the feedback system is functioning well, providing highly relevant, helpful, and accurate insights to users. The data suggests that while there may be small improvements to be made, the system is delivering a strong user experience in terms of feedback quality. These results show the benefit of using GenAI-generated feedback as helpful, accurate and personalised feedback may be obtained quickly.

*4.3 Model Performance for Assessment Personalisation*

The personalised aspect is developed by the GenAI providing a distribution of questions based on user performance. In a nutshell, the mock test should include more questions on poorer-performing topics, so the user can develop more on that topic.

To assess the effectiveness of personalised allocation of questions, 50 simulated users with different performance scores over three topics in the Driving Theory Test were assessed. These performance scores were used to generate a personalised test distribution, which was compared with a distribution by the mathematical model and an expert's allocation. The mathematical model followed the approach of weighting topics based on scores, while the expert derived a more holistic approach using nuance and allocating questions based on a broad understanding of questions. The mean allocation error was used to quantify the average deviation of the LLM's question allocation from these two benchmarks. A lower mean allocation error means our application's allocation is closer to the expert (or mathematical model evaluator).

To calculate the mean allocation error, we calculated the absolute difference between the test allocations from GenAI in our application $G_1, G_2, ..., G_n$ and the benchmark allocations from evaluator $E_1, E_2, ..., E_n$ for each topic $T_1, T_2, ..., T_n$ (in our case, $n = 3$).

$$Mean\ Allocation\ Error\ =\ \frac{1}{n}\sum_{i=1}^{n}|G_i - E_i|$$

We repeated this across 50 users and calculated average errors across the dataset.

The results of the calculation showed a mathematical model benchmark mean error of 1.31 with a standard deviation of 0.96, and the expert benchmark mean error of 1.81, with a standard deviation of 1.18.

The results show that GenAI distribution in our application aligns better with the mathematical model than the expert. It might be explained by the fact that LLMs likely use a fixed inverse-scoring formula (such as allocating more questions to topics with lower scores). This suggests the GenAI effectively replicates performance-based weighting logic.

However, the higher average error of approximately 1.8 against expert benchmarks highlights opportunities for refinement, particularly in capturing nuanced topic relationships.

The higher standard deviation for expert allocation indicates that expert disagreements are wider and the LLM misses subtle prioritisations. Experts may prioritise high-risk topics such as Hazard Perception even if the user score is moderately low, but the LLM might underweight these due to a lack of contextual rules. Experts are likely to allocate questions based on risk criticality, while the GenAI relied solely on performance scores.

In addition, the error distribution showed that 48% of tests had errors >2 against the expert, versus 34% against the model, underscoring gaps in contextual prioritisation.

Although there is room for improvement, the current personalisation is mostly accurate to what an expert would create to help a user. Therefore, this suggests a good use case for GenAI and potentially with more user data a more useful model for real-world cases. By integrating domain-specific rules and topic relationship mapping, the model could achieve more human-like, nuanced allocations.

## 5. Discussion and Future Directions

The results showed that the generated LLM content was largely relevant and helpful, which aligns with the metric-based evaluations conducted by Das *et al.* (2025) and Kanjee *et al.* (2023). However, both these studies use a relatively small sample size. While the model performs well for small sample sizes and in controlled evaluations, larger deployments, as highlighted by Zhao *et al.* (2024), risk introducing inconsistencies and factual inaccuracies, underscoring the need for large-scale, granular validation to ensure robustness.

The system leverages mathematical performance weighting to tailor question distributions. However, our analysis revealed gaps where human expertise outperforms algorithmic logic, particularly in recognising nuanced relationships between topics. For instance, a learner struggling with "Hazard Perception" might also benefit from additional "Road Signs" questions, a connection experts make intuitively, but the current model may miss. These findings echo broader debates in AI education, claiming that pure data-driven approaches often lack the contextual intelligence that teachers apply effortlessly. Ifenthaler *et al.* (2024) suggested that a combination of AI-generated feedback and human intervention creates an optimal learning environment. While incorporating direct teacher feedback was out of scope for this study, there may be a way to provide a teacher interface which helps an educator look through user profiles and adjust their specific settings. The UK Driving Theory test is self-taught so, while not suited for this topic, this would work well in a teaching environment.

Building on these insights and the literature review in Section 2, the following changes may enhance the system efficacy:

*1. Enhancing the Personalisation component*

Currently, the same model is used for all users, this is due to cost concerns and response time. However, deploying user-specific LLM instances trained on individual performance history could enhance personalisation. Adaptive learning platforms like Squirrel AI (Hao, 2019), that leverage diagnostic testing to determine knowledge gaps and personalise learning pathways, could be used to enable dynamic adjustments of question sets based on performance trends, for instance retaining context across mock tests such as persistent weak topics.

*2. Hybrid Feedback Integration*

Though the UK Driving Theory test is self-taught, the platform can benefit from incorporating a teacher interface that allows driving instructors to adjust, as suggested by

Ifenthaler *et al.* (2024). The driving instructors can review flagged content, correct LLM errors, or highlight critical topics to supplement AI output.

### 3. Multimodal Interface

Chan and Hu (2024) emphasised that students engage differently with AI-based learning depending on their individual learning preferences and confidence levels. To address diverse learning preferences, the platform can be enhanced by supporting multiple input formats such as voice input, written response, interactive diagrams (such as drag-and-drop hazard perception exercises), and fill-in-the-blank questions, which may help support different cognitive processes and help diverse learners better engage with the material.

### 4. Gamification Elements

Adding the gamified progress tracking, such as badges for consistent improvement on weak topics, will increase the motivation of learners. Studies such as Bisdas *et al.* (2021) explored the role of AI in motivating students, emphasising that interactive features encourage longer study sessions. To incorporate an engagement-driven design, session persistence was added to allow users to resume tests without losing progress.

## 6. Conclusions

The study aimed to explore the potential intersection between generative AI and the wider education sector through the design and development of an adaptive learning platform. The resulting platform included a web application that employs Gemini 1.5 as an LLM and helps users to revise for their UK Driving Theory Test. Key features included the dynamic generation of questions, personalised feedback, and the construction of mock tests tailored to individual user performance. A series of experiments was conducted to assess the accuracy, relevance, and pedagogical value of the AI-generated content.

The results demonstrated that the GenAI system successfully generated relevant and accurate questions, accompanied by constructive feedback. The personalised mock tests produced by the model were closely aligned with those created by human experts. Quantitative analysis shows strong alignment with established test standards, particularly for structured question formats. Adaptive algorithms adjust content distribution based on individual performance, creating personalised learning pathways that address specific knowledge gaps.

Three key challenges emerged from our findings: innovation, relationships between areas, and scaling. First, while the AI reliably produces conventional questions, it struggles to generate more innovative or open-ended problem formulations. Second, the system's personalisation, though effective at identifying discrete weaknesses, does not fully capture the interconnected nature of driving concepts. Third, scaling the technology while maintaining quality and accommodating diverse learner needs presents a practical implementation issue.

These limitations suggest that future development should focus on enhancing the system's contextual understanding, expanding its capacity for creative problem formulation, and establishing robust quality assurance mechanisms.

Our findings suggest that AI-powered adaptive learning systems have the potential to significantly reshape the landscape of standardised test preparation. The success of this driving theory application provides a model for similar implementations in other domains where structured knowledge assessment is required. As the technology matures, it promises to make personalised, high-quality learning experiences more accessible and effective across educational contexts.